\documentclass[10pt]{article}

\usepackage[margin=1in]{geometry}
\usepackage{times} 
\usepackage{graphicx}
\usepackage{amsmath,amssymb,amsfonts,amsthm}
\usepackage{algorithm}
\usepackage{algpseudocode}
\usepackage{booktabs}
\usepackage{multirow}
\usepackage{url}
\usepackage{hyperref}
\usepackage{enumerate}
\usepackage{enumitem}
\usepackage{listings}
\usepackage{tcolorbox}
\usepackage{subcaption}
\usepackage{mathtools}
\usepackage{cite}
\usepackage{balance} 

\hypersetup{
    colorlinks=true,
    linkcolor=blue,
    filecolor=magenta,
    urlcolor=cyan,
    citecolor=blue
}

\theoremstyle{plain}

\theoremstyle{definition}

\theoremstyle{remark}

\usepackage{geometry}
\usepackage{fancyhdr}
\fancypagestyle{plain}{%
    \fancyhf{}
    \fancyhead[L]{Preprint}
    \fancyhead[R]{Mochizuki et al.}
    \fancyfoot[C]{\thepage}
    
}

\allowdisplaybreaks[4]

\title{%
    \LARGE \textsc{Exposing Citation Vulnerabilities in Generative Engines}
}

\author{
    \textbf{Riku Mochizuki$^{1,2,}$}\thanks{These authors contributed equally to this work }\text{  ,} \textbf{Shusuke Komatsu$^{1,3,*}$}, \textbf{Souta Noguchi$^{1,4}$}, \textbf{Kazuto Ataka$^{4}$}\\ \\
    {\small \textit{$^{1}$QueryLift Inc., Tokyo, Japan}}\\
    {\small \textit{$^{2}$Graduate School of Media and Governance, Keio University, Kanagawa, Japan}}\\
    {\small \textit{$^{3}$Graduate School of Science and Technology, Nara Institute of Science and Technology, Nara, Japan}}\\
    {\small \textit{$^{4}$Faculty of Environment and Information Studies, Keio University, Kanagawa, Japan}}\\
    {\small moz@sfc.keio.ac.jp, \{shusuke.komatsu, souta.noguchi\}@querylift.co.jp, ataka@sfc.keio.ac.jp}
}

\date{}

\begin{document}

\maketitle

\begin{abstract}
We analyze generative engines (GEs) under poisoning attacks from the perspectives of publishers of citations in answers and characterize the attack surface of GEs to poisoning attacks.
GEs integrate two functions: web search and answer generation that cites web pages using large language models.
Because anyone can publish information on the web, GEs are vulnerable to poisoning attacks.
Existing studies of citation evaluation focus on how faithfully answers reflect cited content, leaving unexamined which web sources should be selected as citations to defend against poisoning attacks.
To fill this gap, we introduce evaluation criteria that characterize the attack surface of GEs to poisoning attacks using the publisher attributes of citation contents in answers.
Our criteria classify the publisher attributes of citations to estimate the \emph{content-injection barrier}, which is the difficulty of publishing content with specific publisher authority.
We show a bias in GEs regarding which \emph{content-injection barriers} of contents they preferentially cite and how faithfully they reflect the content in generated answers in political domains in Japan and the United States (U.S.) using our criteria, then reveal the attack surface of GEs and the threat of poisoning attacks.
Our results show that citations from official party websites (primary sources) account for approximately $25\%$--$45\%$ in the U.S. and $60\%$--$65\%$ in Japan, indicating that U.S. political answers have a wide attack surface to poisoning attacks.
We also find that sources with low content-injection barriers are frequently cited yet tend to be poorly reflected in answers.
\end{abstract}
\section{Introduction}\label{sec:introduction}
Large language model (LLM) applications such as GPT~\cite{chatgpt}, Gemini~\cite{gemini}, and Claude~\cite{claude} increasingly incorporate web search functions, providing internet users with new ways of accessing information~\cite{webgpt-arxiv21, geo-apmr+24, zhai2024llm-and-future-of-information-retrieval}. 
Information retrieval has become a primary use case for LLMs, with a survey in September \(2025\) showing that about \(30\%\) of all prompts involve information seeking~\cite{how-people-use-chatgpt-nber25}. 
Systems that perform web searches and generate answers are known as \emph{generative engines} (GEs)~\cite{geo-apmr+24}. 
Web search services are rapidly transitioning from traditional search engines to GE-based systems~\cite{openai-chatgpt-search24, google-gemini-search24, perplexity-comet25, dia-browser25, microsoft-bing-chat23, anthropic-multiagent25}, changing the way users access and interpret web contents.

GEs not only alter the delivery of information but also reshape its nature. 
Whereas conventional search engines direct users to primary information sources such as official documents, GEs provide synthesized answers that summarize and interpret web content. 
LLMs also exhibit bias~\cite{bias-and-fairness-in-large-language-models-cl24, on-the-dangers-of-stochastic-parrots-facc21, holisticevaluationlanguagemodels-arxiv22, bbc-ai-search-engines-2024}, meaning that users receive secondary or higher-order information produced by LLMs rather than directly reading primary sources, similar to human-to-human information delivery~\cite{the-spread-of-true-and-false-news-online-science2018}.
This transformation enhances accessibility but centralizes the selection and interpretation of web content in the GE; moreover, major GEs are black-box systems, introducing vulnerabilities such as inaccuracy of answers.

GEs cite content published on the web by various publishers, including attackers with malicious intent or for profit; therefore, they remain vulnerable to poisoning attacks that inject malicious content into the web and produce attacker-intended answers~\cite{poisonedrag-arxiv24, poisoning-web-scale-training-datasets-is-practical-sp24}. 
When GEs generate answers lacking factuality, the cause is not only LLM hallucinations~\cite{sirens-song-in-the-ai-ocean-coli25, survey-of-hallucination-in-natural-language-generation-acmcsurv23} but also the lack of factuality in the web content obtained through web searches~\cite{retrieval-helps-or-hurts-deeper-arxiv24, poisonedrag-arxiv24, a-comprehensive-survey-on-trustworthy-recommender-systems-arxiv22}.
Web content can be published and modified by any user with any intent~\cite{poisoning-web-scale-training-datasets-is-practical-sp24, poisonedrag-arxiv24}.
Prior studies formalize poisoning attacks that place disinformation and misinformation on the web to generate attacker-intended answers on retrieval from external databases and answer generation systems including GEs because these systems can cite such contents and generate text that reflects them~\cite{poisoning-web-scale-training-datasets-is-practical-sp24, poisonedrag-arxiv24, yu2025safetydevolutionaiagents}.
Specific to GEs, \emph{Generative Engine Optimization} (GEO) presents patterns of which styles of web content GEs prefer to cite in answers~\cite{geo-apmr+24}.
Although GEO is beneficial for primary information providers that seek to increase exposure of their web content as citations, it can also facilitate poisoning attacks.

Existing evaluation criteria aim to assess how faithfully the cited content is reflected in the answers generated by the retrieval and generation systems ~\cite{citeeval-acl25, ragas-eacl24, ares-naacl24, evaluating-verifiability-in-generative-search-engines-emnlp23, enabling-llms-to-generate-text-with-citations-emnlp23, ranking-generated-summaries-by-correctness-acl2019, evaluating-the-factual-consistency-of-abstractive-text-summarization-emnlp20, alignscore-acl23, feqa-acl20}.
These studies analyze textual and semantic consistency between the cited web content and generated answers.
However, these studies alone cannot capture the attack surface of GEs to PoisonedRAG attacks.
This is because these studies targeted systems called \emph{RAG systems}, which retrieve from curated, fixed, and closed external databases to generate answers.
In contrast, GEs treat the web as an external database, which is open to anyone and has dynamism, allowing attackers to easily inject malicious content that GEs can cite.
Due to these differences in characteristics, evaluation criteria proposed for \emph{RAG systems} cannot adequately assess the reliability and safety of GEs.

To address this gap, we introduce novel evaluation criteria that focus on the publisher attributes of citation in answers. 
Our goal is to reveal the bias of GEs in which publisher attributes GEs preferentially cite and how faithfully they reflect the content in generated answers.
First, we classify the publisher attributes of citations in answers following the steps: (1) classification of publishers into primary and secondary information sources, (2) classification of secondary information sources into fine-grained publisher categories using a \emph{LLM-as-a-Judge} method~\cite{survey-llm-as-a-judge-arxiv24, judging-llm-as-a-judge-with-mt-bench-and-chatbot-arena-arxiv23} based on URL and WHOIS data, (3) we classify the category of publishers into \emph{content-injection barriers}, representing the practical difficulty of publishing content with specific publisher authority, specifically into low-, medium-, or high-barrier levels to reveal the attack surface of GEs and the difficulty for poisoning attacks to succeed.
Second, we propose two evaluation criteria: (a) distribution of content-injection barriers in answers and (b) distribution of semantic consistency by content-injection barriers.
Using our criteria, we reveal the bias of GEs in their citation selection and content reflection in answers from the perspective of the authority of publishers, and characterize the attack surface of GEs and the difficulty for poisoning attacks to succeed.

To reveal the threat of poisoning attacks on representative GEs (OpenAI GPT-5, Claude 4 Sonnet, and Gemini Flash 2.0 with search mode enabled), we apply these evaluation criteria to information-retrieval questions in the political domain across Japan and the United States (U.S.), using \(280\) questions and \(4{,}200\) generated answers.
Our results show that citations to official party websites (primary information) constitute approximately \(60\%\)--\(65\%\) of citations in Japan but only \(25\%\)--\(45\%\) in the U.S. and low-barrier sources that can be published with only registration (Reddit, X, personal blog, etc.) account for approximately \(30\%\) of citations in answers.
We further find that citations from low-barrier sources tend to have lower semantic consistency with the actual answer content than medium- and high-barrier sources, yet they still influence GE answers despite their lower semantic consistency.

In summary, our study makes the following contributions:
First, we propose new evaluation criteria with publisher attributes of citations in generated answers by GEs beyond existing evaluation criteria for \emph{RAG systems}.
Second, we find that GEs cite primary sources in \(60\%\)--\(65\%\) of Japanese political answers and \(25\%\)--\(45\%\) of U.S. ones, while low-barrier sources account for approximately \(30\%\) of citations, characterizing the attack surface of GEs and the threat of PoisonedRAG attacks, as GEs tend to cite low-barrier sources.

\section{Generative Engine}
\label{sec:generative_engine}
This section introduces GE and its system model.
GEs are information retrieval systems that integrate web search results from queries transformed from user questions and generate synthesized answers while citing web content using large language models (LLMs)~\cite{geo-apmr+24}.
GEs receive a question from users, perform web searches, and generate answers with LLMs using the search results.

Pranjal et al.\ formalize GEs and provide a system model~\cite{geo-apmr+24}.
A GE is formalized as a function \( f_{GE}^{model} \) that takes a user question \( q_u \) and personalization information \( P_U \) as inputs and generates a textual answer \( r \):
$ f_{GE}^{model} := (q_u, P_U) \rightarrow r $.

GEs consist of two components: content retrieval and answer generation.
Content retrieval collects the information necessary to generate answers from the web.
The user query \( q_u \) is converted by LLMs into multiple queries \( Q^{\prime} = \{q_1, q_2, \cdots, q_n\} \) for the web search.
These queries are sent to a search engine \( SE \), and GEs obtain a set of web sources \( S = \{s_1, s_2, \cdots, s_m\} \) from the search results.
Each result set in $S$ for a web search query \( q_i \) typically consists of the top \( k \) web sources ranked by the search engine's metrics~\cite{poisonedrag-arxiv24, modern-information-retrieval-a-brief-overview-3320, geo-apmr+24}.

Answer generation cites the web sources obtained in the content retrieval phase and generates answers to user queries.
Web sources \( S \) are converted into a summary set \( Sum = \{Sum_1, Sum_2, \cdots, Sum_m\} \) that extracts and summarizes the content of web sources for generating answers to the user query \( q_u \).
LLMs then generate the final answer \( r \) from the summary set \( Sum \) while citing web sources \( S \).
The answer \( r \) consists of \( k \) sentences \( \{l_1, l_2, \cdots, l_k\} \), and each sentence \( l_i \) is associated with a citation set \( C_i \subseteq S, C_i = \{c_1, c_2, \cdots, c_l\} \).
Although a GE ideally has one or more citations for each sentence, there are cases where \( C_i = \emptyset \), meaning there are no citations for the corresponding sentence.

A related technology is Retrieval-Augmented Generation (RAG), which retrieves relevant content from outside of LLMs and incorporates it into LLM answer generation~\cite{internet-augmented-language-models-fewshot-arxiv22, internet-augmented-dialogue-generation-arxiv21, rag-kdd24}.
However, the term \emph{RAG systems} generally refers to systems that retrieve from fixed, curated, and closed external databases defined by developers~\cite{mirage-acl24, lála2023paperqa, almanac-nejmai-2024}.
Note that we distinguish the term \emph{RAG systems} from the term \emph{RAG} in this study. 
\emph{RAG} indicates a technology that retrieves relevant content from outside of LLMs and incorporates it into LLM answer generation, and \emph{RAG systems} are systems that use \emph{RAG} as a core component and target fixed, curated, and closed external databases.
In contrast, while GEs similarly use the web as an external database, the nature of this database differs from that of \emph{RAG systems}.
The web targeted by GEs is open, allowing anyone to freely publish information, and is dynamic, with information being updated as needed.
Therefore, unlike \emph{RAG systems}, GEs use search engines (SE) for retrieval from the external database (web).
In our study, we focus on the differences in external database characteristics between GEs and \emph{RAG systems}.

\section{Related Works and Motivation}
This section introduces poisoning attacks on \emph{RAG systems} and GEs, seminal evaluation criteria for \emph{RAG systems}, and the gap between GEs and \emph{RAG systems} that motivates our evaluation criteria.

\label{sec:related}
\subsection{Vulnerabilities of GEs}
We introduce a poisoning attack in GEs.
Zou et al.~\cite{poisonedrag-arxiv24} reveal and formalize an attack method called \emph{PoisonedRAG} that exploits \emph{RAG systems} to generate attacker-intended answers by placing malicious content into external databases.
Even if attackers can inject a small volume of malicious text into external databases, \emph{RAG systems} can produce attacker-intended answers for specific questions.
This malicious content includes text that attackers want displayed as answers to specific questions, especially false information such as ``\textit{OpenAI's CEO is Tim Cook.}''
PoisonedRAG attacks target closed-ended questions like ``\textit{Who is the CEO of OpenAI?}'' rather than open-ended questions like ``\textit{What are the latest trends in AI?}''~\cite{poisonedrag-arxiv24}.

In fact, several studies support the effectiveness of this attack.
Chen et al.~\cite{chen2022rich} show that in \emph{RAG systems}, when content containing two different claims (e.g., \textit{OpenAI's CEO is Tim Cook.} and \textit{OpenAI's CEO is Sam Altman.}) is placed and answers are generated for each claim, the answers may include each respective claim.
Wang et al.~\cite{wang2025ragconflictevidence} and Liu et al.~\cite{liu2025conflicts} demonstrate that even when there is an imbalance in the volume of supporting documents for each claim, \emph{RAG systems} tend to favor the answer with more supporting evidence, yet the underrepresented claim can still appear in responses.
These studies support the effectiveness of PoisonedRAG attacks and demonstrate that RAG may cite malicious content even when attackers place only a small volume of it.

To succeed in a poisoning attack, the malicious content must satisfy two conditions: \emph{Retrieval Condition} and \emph{Generation Condition}.
Retrieval Condition means that malicious content (such as text or HTML placed by attackers) is selected as top-\( k \) relevant content through retrieval tasks for target questions.
In GEs, Retrieval Condition means that content on the web is retrieved as web pages ranked highly in search engine results by SE.
Generation Condition means that the content is used as reasoning knowledge in the LLM and reflected in the generated answers.
Aggarwal et al.~\cite{geo-apmr+24} propose \emph{generative engine optimization} (GEO), a method that optimizes website text structure and content from the perspective of Generation Condition to increase the exposure of web page content as citations in answers when the website passes the Retrieval Conditions.
GEO theoretically supports that GEs are vulnerable to poisoning attacks.

\subsection{Citation Evaluation}
\label{subsec:citation_evaluation}
We first introduce seminal studies that aim to evaluate \emph{RAG systems}.
These studies highlight that in ideal \emph{RAG systems}, all claimed sentences in an answer output from the \emph{RAG system} are supported by external sources from databases as \emph{citations} (citation coverage) and the sentences with citations accurately and faithfully reflect the claims (reflection accuracy).
To analyze them, numerous studies evaluate citation coverage and reflection accuracy, specifically focusing on \emph{RAG systems}~\cite{citeeval-acl25, ragas-eacl24, ares-naacl24, evaluating-verifiability-in-generative-search-engines-emnlp23, enabling-llms-to-generate-text-with-citations-emnlp23, ranking-generated-summaries-by-correctness-acl2019, evaluating-the-factual-consistency-of-abstractive-text-summarization-emnlp20, alignscore-acl23, feqa-acl20}.

A core component of the analysis of reflection accuracy is faithfulness evaluation metrics~\cite{finegrainedcitationevaluationgenerated-arxiv24}.
Faithfulness evaluation metrics are typically classified into three methods: entailment-based~\cite{ranking-generated-summaries-by-correctness-acl2019, true-reevaluating-factual-consistency-arxiv22, evaluating-the-factual-consistency-of-abstractive-text-summarization-emnlp20, alignscore-acl23}, similarity-based~\cite{bertscore-arxiv20, bartscore-neurips21}, and QA-based~\cite{feqa-acl20}.
Similarity-based methods are most appropriate for calculating reflection accuracy.
Similarity-based methods quantitatively measure semantic similarity between two texts and rely on neural encoder models.
Zhang et al.\ show that similarity-based methods achieve higher precision than entailment-based and QA-based methods in tasks evaluating consistency between long documents and their summaries~\cite{finegrainedcitationevaluationgenerated-arxiv24}.
This approach aligns with GE behavior of summarizing web page content retrieved from search results and generating answers with citations.
Citation evaluation applies these faithfulness evaluation methods to each sentence \( l_i \) of generated answer \( r \) with citations \( C_i \), and evaluates reflection accuracy for the entire answer \( r \).

\subsection{Issue and Research Motivation}
Existing seminal studies that focus on citation evaluation to assess reliability and safety of \emph{RAG systems} cannot capture the attack surface to poisoning attacks.
Conventional evaluation criteria for \emph{RAG systems} analyze citation coverage and reflection accuracy of retrieved content from external databases; however, they do not examine which publishers of the content are selected as citations and assess the authority and trustworthiness of those sources.
Thus, a gap remains between the seminal studies on \emph{RAG systems} and their applicability to GEs.
Poisoning attack studies in \emph{RAG systems} focus on how much malicious content attackers need to inject into external databases to succeed in the poisoning attack.
In GEs, since they retrieve information from the open web where attackers can inject malicious content more easily than in \emph{RAG systems}, the primary concern should shift to whether GEs retrieve even a small volume of malicious content injected by attackers during web search, which would result in a successful poisoning attack as discussed above.

To fill this gap, we propose novel evaluation criteria using publisher attributes of citations in generated answers that can assess the reliability and safety of GEs where poisoning attacks can be a threat.
Our goal is to provide evaluation criteria for assessing the attack surface to poisoning attacks from generated answers by GEs, not to demonstrate successful attacks themselves or to determine whether malicious content has actually been injected; rather, we aim to provide evaluation criteria for assessing the potential effectiveness of poisoning attacks based on publisher attributes of citations in observed actual answers and to reveal the attack surface of GEs and the difficulty for poisoning attacks to succeed.

\section{Methodology}
\label{sec:methodology}

\begin{figure*}[htbp]
    \centering
    \includegraphics[width=1.0\textwidth]{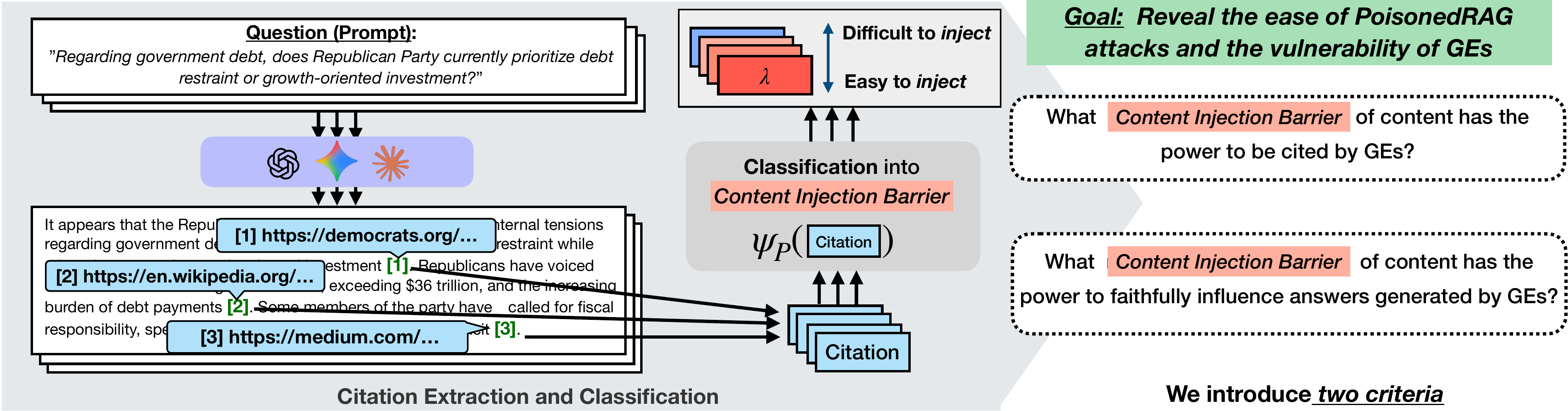}
    \caption{Our research goals and methodology. We aim to characterize the attack surface of GEs to poisoning attacks by analyzing the publisher attributes of citations as \emph{content-injection barriers} in generated answers.}
    \label{fig:overview}
\end{figure*}

We introduce novel evaluation criteria that focus on the publisher attributes of citations in answers to characterize the attack surface of GEs to poisoning attacks.

Our study aims to reveal two types of bias and define research questions:
\emph{R-1}: Which publisher's content GEs preferentially cite and reveal bias of citation selection from the perspective of publishers?
\emph{R-2}: How publisher's content has the power to influence the content of generated answers faithfully?
By providing evaluation criteria that can answer these questions, we enable users and communities to assess whether vulnerable behaviors under PoisonedRAG attacks exist in GEs according to the nature and target of information retrieval, thereby contributing to the monitoring and maintenance of the safety of GEs.

To reveal these biases, we propose a component to classify the publisher attributes of citations in answers.
In this component, we define the difficulty of publishing content on the web with specific publisher attributes as the \emph{content-injection barrier} and categorize publishers of citations based on these barriers into low-, medium-, and high-barrier categories.
Using this barrier, we can reveal which level of publisher authority attackers must achieve to successfully inject malicious content and influence GE answers.
By clustering citations based on content-injection barriers and analyzing GE behavior by the barriers, we can assess the practical difficulty and threat of poisoning attacks.

Finally, we introduce a citation reflection component to reveal the bias of citation reflection from the perspective of publisher attributes.
This component quantifies the influence of each content of citations on answer generation by using similarity-based methods.
We show the overview of our goals and proposed methodology in Figure~\ref{fig:overview}.

\subsection{Citation Classification}
\label{subsec:citation_classification}
The first component classifies citations in answers into content-injection barriers.
We introduce a classifier \( \psi_{\mathcal{P}}(c) \) that assigns each citation \( c \) to an appropriate \emph{content-injection barrier} for each target political party \( \mathcal{P} \).
In our experiments in Section~\ref{sec:experiment}, we target the questions in the political domain; thus, in explanation, we denote \( \mathcal{P} \) as the target political party.
We divide the classification into three components: primary information identification, secondary information category classification, and content-injection barrier classification.
We perform multi-stage classification to ensure interpretability of the clustering results.
Because content-injection barriers are abstract high-level categories, we ensure the validity and interpretability of this categorization by first classifying publishers into primary and secondary information sources, then categorizing secondary sources into fine-grained publisher categories, and finally deriving content-injection barriers from these categories.

\textbf{Primary Information Identification.}
We identify primary information sources that are directly relevant to specific domains.
First, we define the target domain set \( \mathcal{D}_{\mathcal{P}} \) (e.g., $\mathcal{P}$ is the U.S. Democratic Party, \( \mathcal{D}_{\mathcal{P}} = \{\text{democrats.org, democrats.gov, democrats.io}\} \)).
Let \( d_c = \text{domain}(c) \) be a function extracting the domain from each citation \( c \in C \).
We check whether \( \mathcal{D}_{\mathcal{P}} \) includes \( d_c \).
When \( d_c \) is included in \( \mathcal{D}_{\mathcal{P}} \), we classify the citation as a primary information source.
Here, \( \mu_{\mathcal{P}}(c) \) outputs 1 if \( d_c \) is included in \( \mathcal{D}_{\mathcal{P}} \) and \(\bot\) otherwise.

\textbf{Secondary Information Category Classification.}
We categorize secondary information into fine-grained publisher categories, which serve as intermediate categories for calculating content-injection barriers.
We prepare publisher category set \\ \( \Pi = \{\pi_1, \pi_2, \cdots, \pi_p\} \). 
Here, \( \pi(c) \) maps citation \( c \) to category \( \pi; i \in \{1, 2, \cdots, p\} \).
In our experiments in Section~\ref{sec:experiment}, we use labels such as ``Party'', ``Media'', ``Platform'', ``Owned'', ``Academia'', and ``Non-media-industry''.

To build the function \( \pi(c) \), we adopt a hybrid strategy combining automatic category classification using LLM-as-a-Judge~\cite{survey-llm-as-a-judge-arxiv24,judging-llm-as-a-judge-with-mt-bench-and-chatbot-arena-arxiv23} and manual category classification.
We prepare two GE models (ideally two models from different providers such as GPT and Gemini with web search mode enabled) and use them to identify publishers from domains and WHOIS information.
Combining multiple GE models can reduce potential biases of single GE models and increase classifier accuracy~\cite{judging-llm-as-a-judge-with-mt-bench-and-chatbot-arena-arxiv23}.
When classification results from the two GE models agree, we adopt that result directly; when they disagree, we perform manual classification.
In our experiments, the authors make the final determination based on company information from the website and domain registration data.
We employ zero-shot classification~\cite{radford2019language} following Kostina et al.~\cite{kostina2025largelanguagemodelstext}, who demonstrate high performance without training examples.
For each citation \( c \in C \), the prompt receives two inputs: \( \text{url}(c) \) extracts the complete URL indicating the publisher's domain and path, and \( \text{whois}(c) \) retrieves domain registration data including ownership and organizational information.
We provide the complete classification prompt template and the domain-to-category mapping table used in our experiments in our repository~\cite{QL_SIGIR_RERO}.

Finally, we derive content-injection barriers from these categories.
Here, \( \lambda_{\mathcal{P}}(\pi) \) maps publisher category \( \pi \) to content-injection barrier \( \Lambda = \{\lambda_1, \lambda_2, \cdots, \lambda_q\} \). 
We define five content-injection barriers \( \Lambda \): \emph{Primary Sources}, \emph{Opponent Sources}, \emph{Low-Barrier Sources}, \emph{Medium-Barrier Sources}, and \emph{High-Barrier Sources}.
\emph{Primary Sources} are domains owned by the party referenced in the question. If \( \mu_{\mathcal{P}}(c) = 1 \), then \( c \) is classified as a primary information source.
\emph{Opponent Sources} are domains operated by rival parties. If \( \mu_{\mathcal{P}}(c) = \bot \) and \( \pi(c) \) is ``Party'' in our experiment, then \( c \) is classified as an opponent information source.
\emph{Low-Barrier Sources} are domains where users or owners can freely publish or edit content (``Platform'' and ``Owned'' in our experiments).
\emph{Medium-Barrier Sources} are domains owned by organizations or companies with editorial processes where journalism bias or interests may appear (``Media'' and ``Non-media-industry'' in our experiments).
\emph{High-Barrier Sources} are domains where authors are required to remain neutral and objective and manipulation is difficult (``Academia'' and ``Government'' in our experiments).

\subsection{Citation Reflection}
\label{subsec:citation_reflection}
The second component measures the semantic consistency of citations and answers and calculates the citation reflection power for each citation content.
We introduce a classifier \( \tau(c, r) \) that classifies each citation \( c \) and answer \( r \) into an appropriate \emph{similarity band} \( T = \{\tau_1, \tau_2, \cdots, \tau_b\} \).
We decompose the measurement into two steps:

\textbf{Decomposing Answers and Citations into Sets of Sentences.}
First, we decompose answer \( r \) into sentences \( S = \{l_1, l_2, \ldots, l_k\} \) and each citation \( c_i \in C \) into sentences \( S_{c_i} = \{s_{i,1}, s_{i,2}, \ldots, s_{i,n_i}\} \).
Because some sentences of answer \( r \) have citation labels attached and the semantics of each answer sentence is independent, we need to decompose answer \( r \) into sentences to evaluate similarity accurately.

\textbf{Similarity Measurement.}
We measure textual and semantic similarity between each answer sentence \( \{l_1, l_2, \ldots, l_k\} \) and each citation sentence \( S_{c_i} = \{s_{i,1}, s_{i,2}, \ldots, s_{i,n_i}\} \) for each citation \( c_i \in C \).
To measure similarity between two sentences, we introduce the calculation of similarity function \( \text{sim}(x, y) \) between two sentences \( x \) and \( y \).
The function \( \text{sim}(x, y) \) returns a similarity score between the two sentences \( x \) and \( y \) as a value in the range of \( -1 \) to \( 1 \).
Here, a similarity value closer to \(-1\) indicates lower semantic similarity, whereas a value closer to \(1\) indicates higher semantic similarity.
To calculate the similarity, we employ Sentence-BERT~\cite{sentence-bert-arxiv19}, which converts each sentence into dense embedding vectors and computes their semantic similarity through cosine similarity.
In our experiments in Section~\ref{sec:experiment}, we utilize the pre-trained Sentence-BERT model ``stsb-xlm-r-multilingual''~\cite{sentence-bert-arxiv19}, which has been trained on multilingual semantic textual similarity tasks.

\textbf{Maximum Similarity Identification.}
GEs often use only some sentences within long citation texts to generate particular answer sentences, thus we identify the maximum similarity between the answer sentence set \( S \) and citation \( c_i \) to reveal the reflection power of a citation in answers.
Finding the maximum similarity identifies the most relevant part of the citation text to the answer content.
In detail, we calculate the maximum similarity as \( \text{sim}_{\text{max}}(S, c_i) = \max_{j \in \{1,\ldots,k\}, m \in \{1,\ldots,n_i\}} \text{sim}(l_j \in S, s_{i,m} \in S_{c_i}) \).
In calculating this maximum similarity, we compute the similarity between each answer sentence \( l_j \in S \) and all sentences \( s_{i,m} \) within citation \( c_i \) and adopt the highest similarity value across all combinations.

Finally, we categorize the maximum similarity scores \( \text{sim}_{\text{max}}(S, c_i) \) into three \emph{similarity bands} \( T \) to reveal the reflection power of a citation in answers.
We introduce a $\gamma(\text{sim}_{\text{max}}(S, c_i))$ function that maps the maximum similarity score to the similarity band:
\emph{High} similarity (scores in \([0.9, 1.0]\)) indicates strong semantic alignment between answer and citation sentences, suggesting direct reflection of the citation source in the answer.
\emph{Mid} similarity (scores in \([0.8, 0.9)\)) represents moderate semantic overlap, indicating partial citation contribution.
\emph{Low} similarity (scores in \([-1.0, 0.8)\)) reflects weak semantic connection, suggesting minimal reflection of the citation source in the answer.
We employ these thresholds because \(0.8\) has been used as a threshold for detecting paraphrased sentences~\cite{kabir2024banglaembed} and \(0.9\) for classifying similar sentences~\cite{tumre2025improved} in prior studies.


\section{Experiment}
\label{sec:experiment}
This section presents citation patterns in representative and widely used GEs within the political domains of Japan and the United States using the methodology proposed in Section~\ref{sec:methodology}.
The purpose of this experiment is twofold: (1) to reveal the potential attack surface of GEs from the perspective of PoisonedRAG attacks in representative and widely used GEs within the political domain, which is critical because these GEs can significantly influence citizens' decision-making in elections, and (2) to accelerate research addressing these risks by sharing the publisher attribute perspective with the research community.
Our experimental goal is not to demonstrate effectiveness across exhaustive domains, but rather to illustrate the attack surface and threats that exist in the GEs from the perspective of publisher attributes.

\begin{figure*}[htbp]
\begin{subfigure}[b]{0.32\textwidth}
    \centering
    \includegraphics[width=\textwidth]{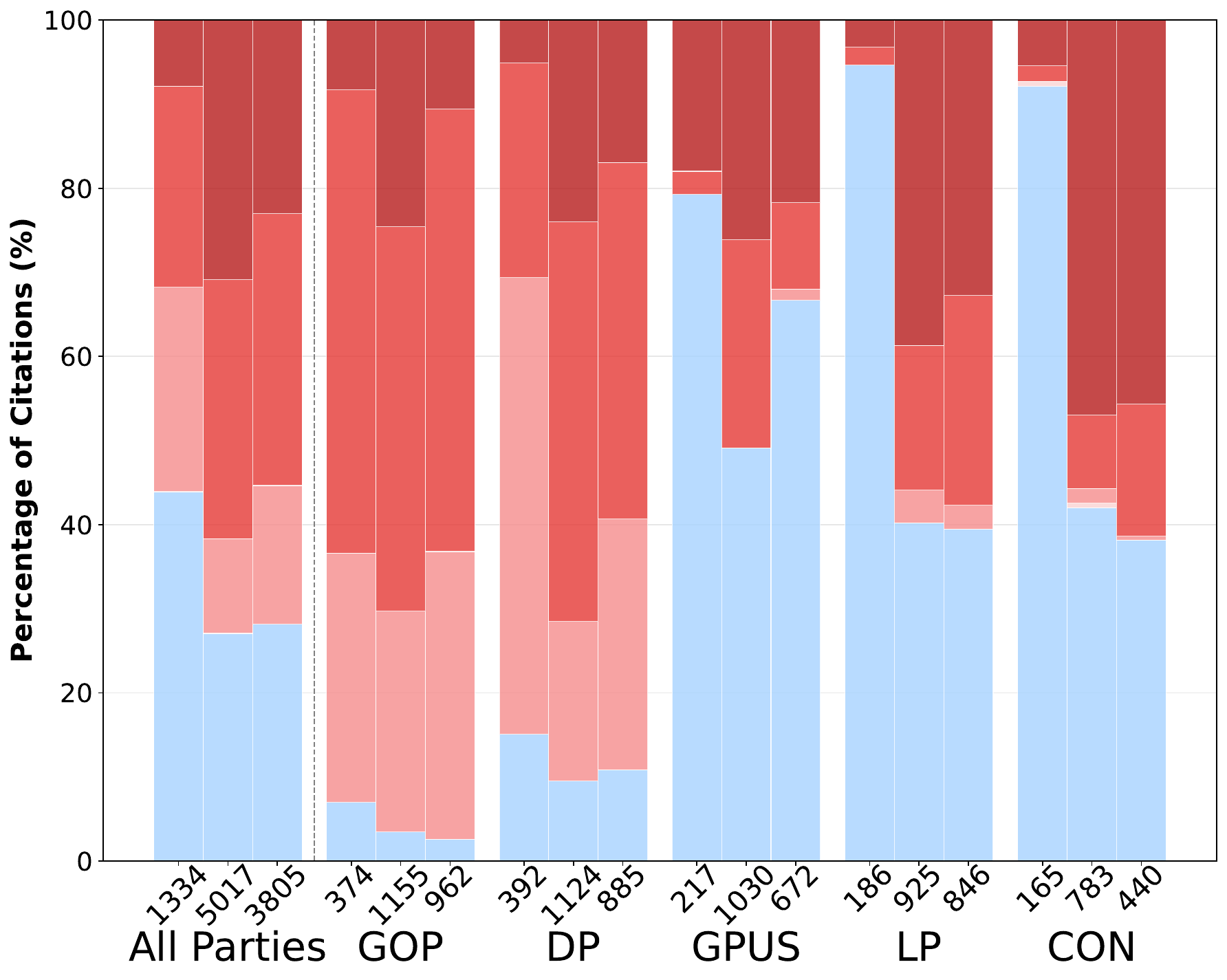}
    \caption{Citation Distributions (U.S.)}
    \label{fig:all_en}  
\end{subfigure}
\begin{subfigure}[b]{0.64\textwidth}
    \centering
    \includegraphics[width=\textwidth]{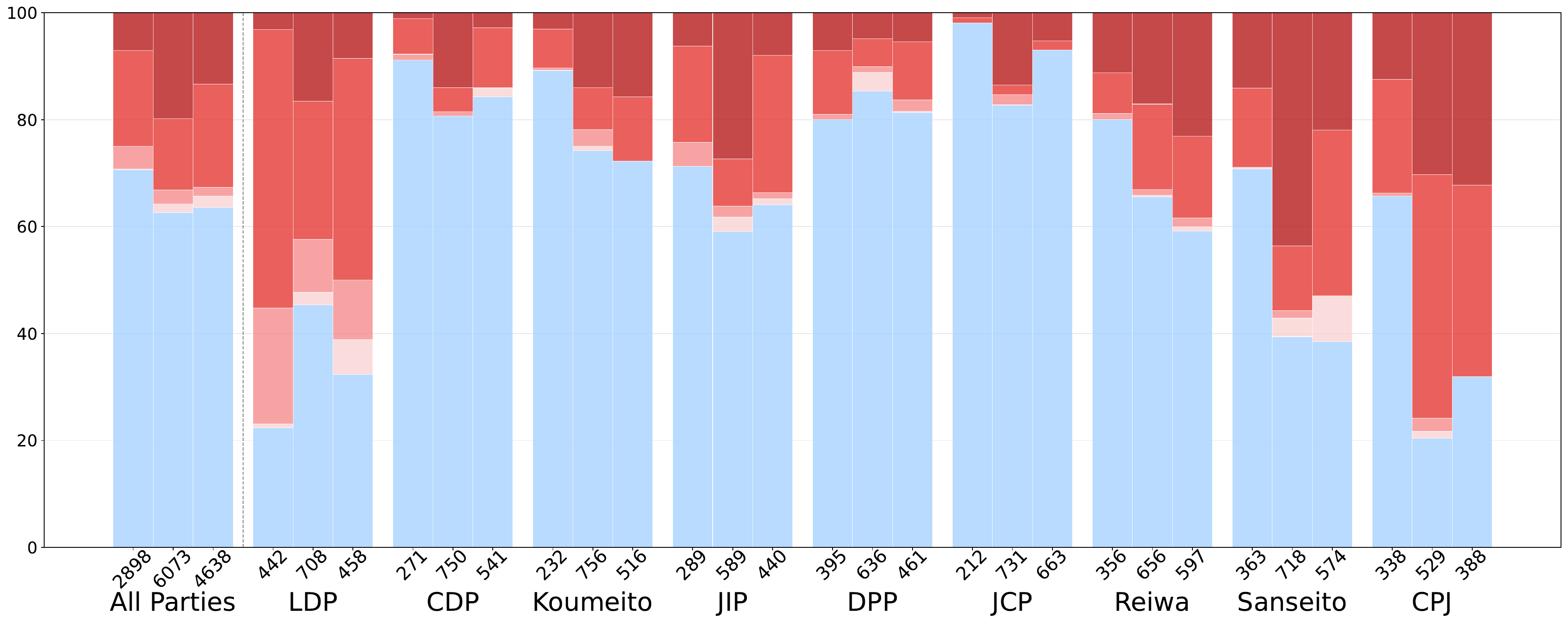}
    \caption{Citation Distributions (Japan)}
    \label{fig:all_ja}
\end{subfigure}
\begin{subfigure}[b]{0.99\textwidth}
    \centering
    \includegraphics[width=\textwidth]{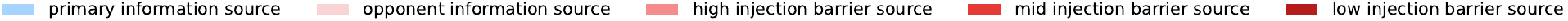}
\end{subfigure}
\caption{Distribution of citation sources by party in the US (a) and Japan (b): Each stacked bar shows the proportion of cited sources (primary, opponent, and secondary information sources categorized by attack cost) for responses generated by different APIs.
For each party, the three adjacent bars correspond to results from OpenAI (left), Gemini (middle), and Claude (right).
The three bars on the far right of each panel show aggregated results across all parties. The ``opponent source'' in All Parties chart reffers to the citation sources from parties that are not the target party for each question. The numbers under each chart shows the amout of total citation constructing it.}
\label{fig:party_wise}
\end{figure*}

\subsection{Experimental Setup}
We focus on the political domain because the flow of information from political parties to citizens constitutes a cornerstone of democratic electoral processes~\cite{the-trust-gap-young-peoples-tactics-for-assessing-the-reliability-of-political-news-ijpp22, kovach2007elements}.
Contemporary elections depend critically on digital information dissemination, and citizens increasingly rely on GEs to gather information that shapes their voting decisions.
GEs can significantly influence citizens' decision-making, rendering the threat of PoisonedRAG attacks particularly severe in this domain.
We emphasize that our experiment aims to measure actual citation patterns in the political domain rather than prescribe ideal proportions; we defer discussion of desirable citation balance for GEs to Section~\ref{sec:discussion}.

We design political questions about topics common to both countries and generate answers to those questions.
We prepare ten policy questions and ten ideology questions, totaling twenty questions.
All questions are closed-ended and include a party name according to PoisonedRAGs study (e.g, ``Regarding government debt, does {PARTY} currently prioritize debt restraint or growth-oriented investment?'').
We show all questions in our repository~\cite{QL_SIGIR_RERO}.
The scale of our question set is comparable to seminal evaluation studies, which typically use \(50\)--\(100\) questions (e.g., Guan et al.~\cite{guan2024language}; Wang et al.~\cite{wang2024factcheck}).

We clarify that our questions are designed to elicit policy positions without bias toward specific parties.
Our questions are designed to be politically neutral, focusing on whether GEs cite contents that articulate positions on the target topics rather than favoring any particular party; consequently, our findings reflect citation patterns rather than biases introduced by question design.
The questions are prepared in two languages: Japanese for Japanese politics and English for U.S. politics.
Answers to Japanese questions are generated in Japanese, and answers to U.S. questions are generated in English.

We target political parties that satisfy each country's requirements for national political parties as primary information sources.
Specifically, in Japan we target nine parties: ``Liberal Democratic Party (LDP)'', ``Constitutional Democratic Party of Japan (CDP)'', ``Komeito'', ``Japan Innovation Party (JIP)'', ``Democratic Party for the People (DPP)'', ``Japanese Communist Party (JCP)'', ``Reiwa Shinsengumi (Reiwa)'', ``Sanseito'', and ``Conservative Party of Japan (CPJ)''; in the U.S. we target five parties: ``Republican Party (GOP)'', ``Democratic Party (DP)'', ``Green Party (GPUS)'', ``Libertarian Party (LP)'', and ``Constitution Party (COP)''.
We prepare twenty questions for each party, resulting in \(180\) questions for Japan and \(100\) questions for the U.S. respectively.

We employ three GE models for answer generation: OpenAI GPT-5~\cite{chatgpt}, Claude Sonnet \(4\) (claude-4-sonnet-20250514)~\cite{claude}, and Gemini Flash \(2.0\) (gemini-2.0-flash)~\cite{gemini}, the most advanced publicly available models with web search mode enabled.
We clarify why we focus on closed GE models in this study.
According to surveys~\cite{menlo-ventures2025stategenaienterprise, a16z2025stateconsumerai}, closed models dominate the LLM market in both enterprise and consumer segments, with Claude, OpenAI, and Gemini together accounting for 88\% of spending share, while open-source models account for only 11\%.
These models are developed in the U.S., but they dominate usage in Japan as well.
According to GMO Research \& AI~\cite{gmo-research2025japanai}, Japanese users predominantly rely on closed models such as GPT, Gemini, and Copilot, all of which are U.S.-developed products.
While Japan has developed numerous LLM models, to our knowledge, no Japanese GE models are publicly available.
Furthermore, widely deployed GEs such as Google AI Overview, which integrates Google Search by default, rely on the closed model Gemini~\cite{stein2025expandingaioverviewsaimode, reid2025aisearchbeyondinformation}; consequently, users routinely encounter generated answers from closed GE models when conducting web searches, underscoring the central role these models play in contemporary information retrieval.
Note that we do not limit our evaluation criteria to closed models because our evaluation criteria do not use the internal reasoning states of GEs but only use input and output of GEs (question and answer).
To provide reproducibility, the question-answer datasets are available in our repository~\cite{QL_SIGIR_RERO}.

In our experiments, the temperature of each GE model cannot be set because APIs do not allow setting temperature for LLMs with search mode enabled; thus, the cited web sources and textual content of answers vary across generations.
To capture answer variation, we ask each question five times to suppress bias from a single answer.
The answers are obtained on September \(4\), \(2025\).

\subsection{Distribution of Content-injection Barriers in Answers}
\label{sec:api_response}
This sub-experiment answers which publisher attributes from the perspective of content-injection barriers GEs preferentially cite (\emph{R-1}).
We calculate the proportion of each content-injection barrier using the citation classification method in Section~\ref{subsec:citation_classification} as \( P_{\lambda_i} = |C^{\lambda_i}|/|C| \), where \( |C^{\lambda_i}| \) is the number of citations in content-injection barrier \( \lambda_i \) and \( |C| \) is the total number of citations.

We show the aggregated results in Figures~\ref{fig:all_en}--\ref{fig:all_ja}.
These figures present stacked proportions by party and model.
Each group of three stacked bar graphs shows the proportions of publisher attribute categories of citation web sources in the answer on the horizontal axis.
For each party, the stacks are ordered left to right as OpenAI, Gemini, and Claude, respectively.
The three graphs at the left edge of each figure show the combined results of all citations across parties as a baseline for each model.

We first analyze the overall trends in citation patterns.
Overall, for Japanese questions (Japanese parties), all three models show a high proportion of target-party primary information sources, accounting for about \(60\%\): OpenAI \(63.0\%\), Gemini \(60.9\%\), and Claude \(60.9\%\).
In contrast, for U.S. questions (U.S. parties), the proportion of primary information sources decreases (OpenAI \(43.9\%\), Gemini \(27.0\%\), Claude \(28.1\%\)), and the proportion of secondary information sources increases.
We observe a large structural difference: primary source dependence in the Japanese setting and external source dependence in the U.S. setting.

We observe distinct differences in citation patterns across the three models.
Gemini consistently exhibits strong platform dependence: low-barrier sources contribute \(19.9\%\) in Japanese questions and \(30.9\%\) in U.S. questions.
Claude tends to use media-related sources the most: medium-barrier sources contribute \(19.3\%\) in Japanese questions and \(32.3\%\) in U.S. questions.
OpenAI cites categories in a more balanced manner: medium-barrier sources \(17.7\%\) and low-barrier sources \(7.2\%\) in Japanese questions, whereas high-barrier sources \(24.3\%\) and medium-barrier sources \(23.9\%\) in U.S. prompts.
Gemini is platform-leaning, Claude is media-leaning, and OpenAI is distributed.

We identify significant differences in citation patterns between Japanese and U.S. contexts.
With Japanese questions, all three models converge to around a \(60\%\) share for primary information sources, and opponents and secondary information sources are used in a supplementary role.
With U.S. questions, the share of primary information sources falls to the \(20\%\)--\(40\%\) range, and the shortfall is supplemented by different opponent and secondary information sources depending on the model.
Concretely, in the U.S., Gemini emphasizes platform sources (\(30.9\%\)), Claude emphasizes media sources (\(32.3\%\)), and OpenAI emphasizes high-barrier sources (\(24.3\%\)), indicating that the direction of external dependence diverges substantially by country and language.

We examine citation patterns for individual U.S. political parties.
For both the Republican and Democratic Parties, the two major parties in the U.S., the proportion of opponents and secondary information sources is high across models, with medium- and high-barrier sources as the main components.
Except for questions about the Democratic Party with OpenAI, the two parties often cite low- and medium-barrier sources.
In contrast, the Green Party frequently cites low-barrier sources.
Other parties tend to cite official party sources more, though the volume varies across models and parties.
For the Green Party, OpenAI and Claude have high shares of primary information sources (\(80\%\) and \(70\%\), respectively), whereas for Gemini roughly \(10\%\) of that share is replaced by medium-barrier sources.
For the Libertarian Party and the Constitution Party, similar tendencies are observed: only OpenAI shows a remarkably high share (greater than \(90\%\)) from primary information sources, whereas other models increase dependence on low-barrier sources to around \(40\%\).

We analyze citation patterns for individual Japanese political parties.
The current ruling party, the Liberal Democratic Party, shows a mitigated concentration on primary information sources compared with other parties.
Meanwhile, citations from high-barrier sources, which are opponents and secondary information sources with high barriers, increase; nevertheless, compared with questions about other parties, citations from high-barrier opponents and secondary information sources remain relatively high.
For the Constitutional Democratic Party of Japan, Komeito, and the Japan Innovation Party, primary information sources generally account for \(70\%\)--\(90\%\), and secondary information sources remain supplementary.
The Japanese Communist Party exhibits particularly high dependence on the party's official domain, with answer justifications concentrating in intra-party information.
For Sanseito and the Conservative Party of Japan, the shares from platforms and media increase relatively in Gemini and Claude, indicating a structure in which the shortage of party primary information sources is supplemented by opponents and secondary information sources.
Looking at the results for the Liberal Democratic Party in Japan and the Republican Party and Democratic Party in the U.S., questions about the major parties in their countries tend to have low citations from primary information sources.
This could be because those parties are major parties in their countries, and there may be influence from the vast number of secondary information sources, especially in government and media domains.

\begin{figure*}[htbp]
    \centering
    \vspace{1em}
    \begin{subfigure}[t]{0.32\textwidth}
        \centering
        \includegraphics[width=\textwidth]{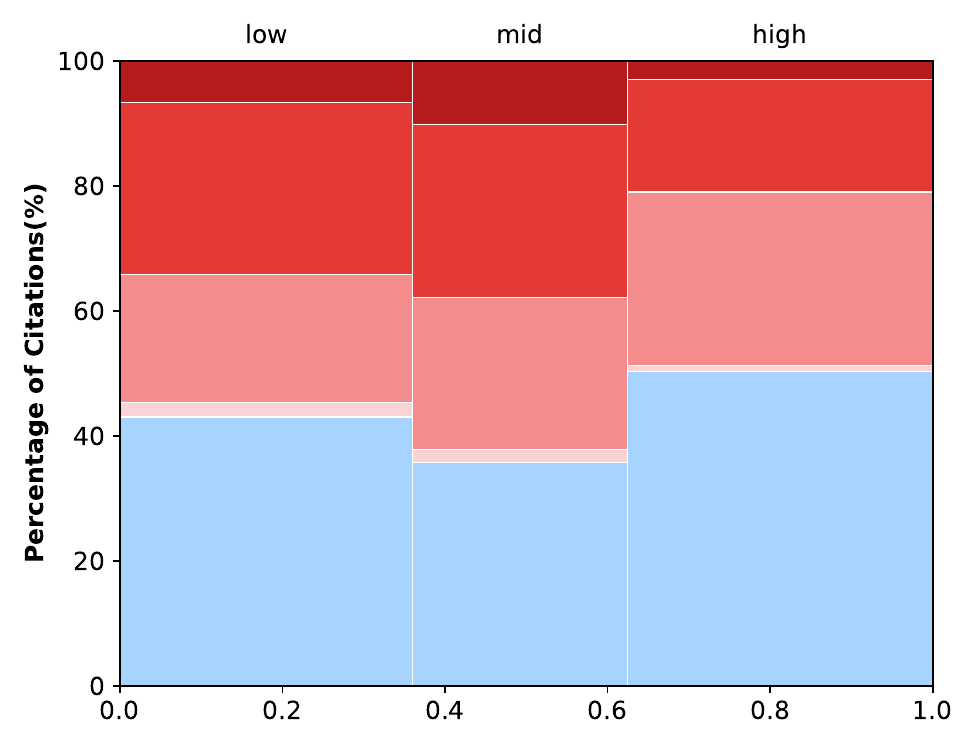}
        \caption{Distributions in OpenAI (U.S.)}
        \label{fig:citation_coverage_en_openai}
    \end{subfigure}
    \begin{subfigure}[t]{0.32\textwidth}
        \centering
        \includegraphics[width=\textwidth]{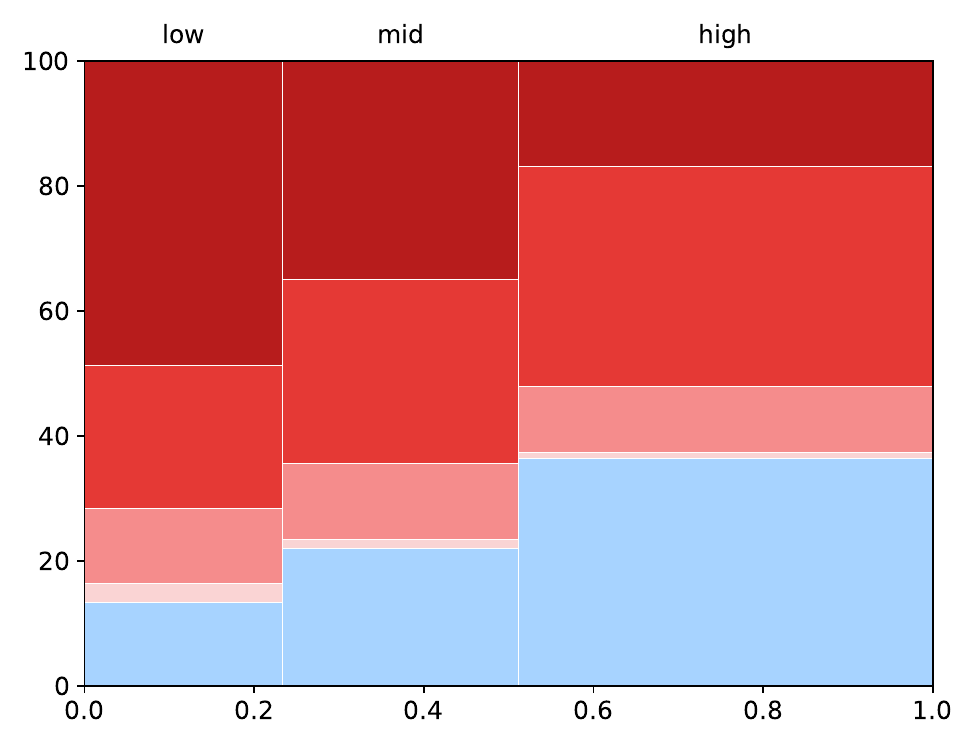}
        \caption{Distributions in Gemini (U.S.)}
        \label{fig:citation_coverage_en_gemini}
    \end{subfigure}
    \begin{subfigure}[t]{0.32\textwidth}
        \centering
        \includegraphics[width=\textwidth]{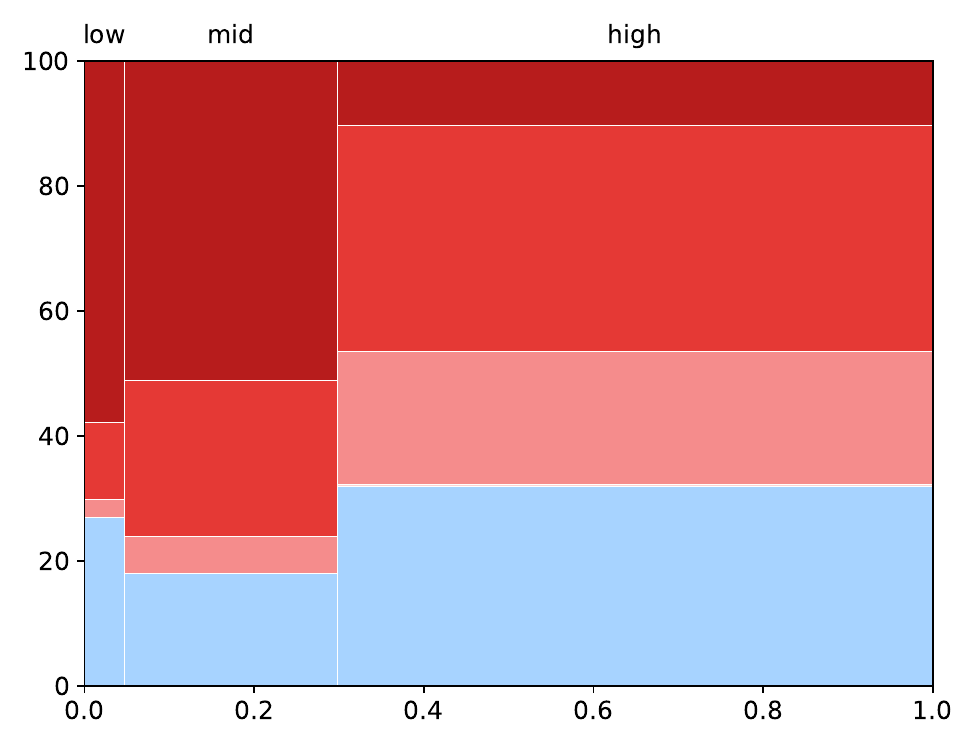}
        \caption{Distributions in Claude (U.S.)}
        \label{fig:citation_coverage_en_claude}
    \end{subfigure}
    \begin{subfigure}[t]{0.32\textwidth}
        \centering
        \includegraphics[width=\textwidth]{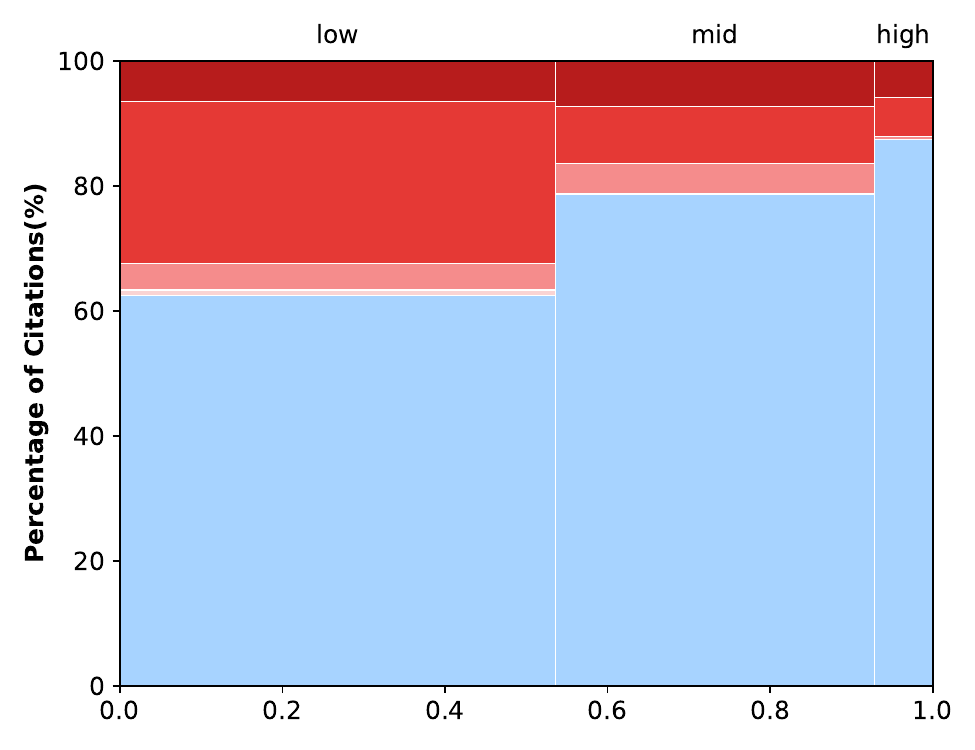}
        \caption{Distributions in OpenAI (Japan)}
        \label{fig:citation_coverage_ja_openai}
    \end{subfigure}
    \begin{subfigure}[t]{0.32\textwidth}
        \centering
        \includegraphics[width=\textwidth]{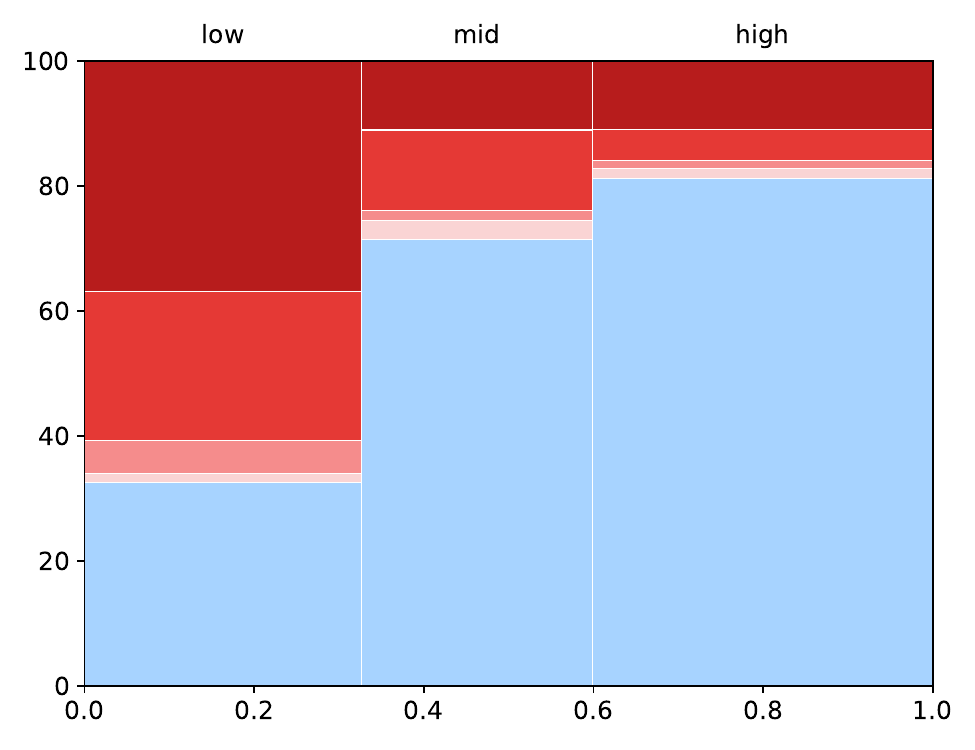}
        \caption{Distributions in Gemini (Japan)}
        \label{fig:citation_coverage_ja_gemini}
    \end{subfigure}
    \begin{subfigure}[t]{0.32\textwidth}
        \centering
        \includegraphics[width=\textwidth]{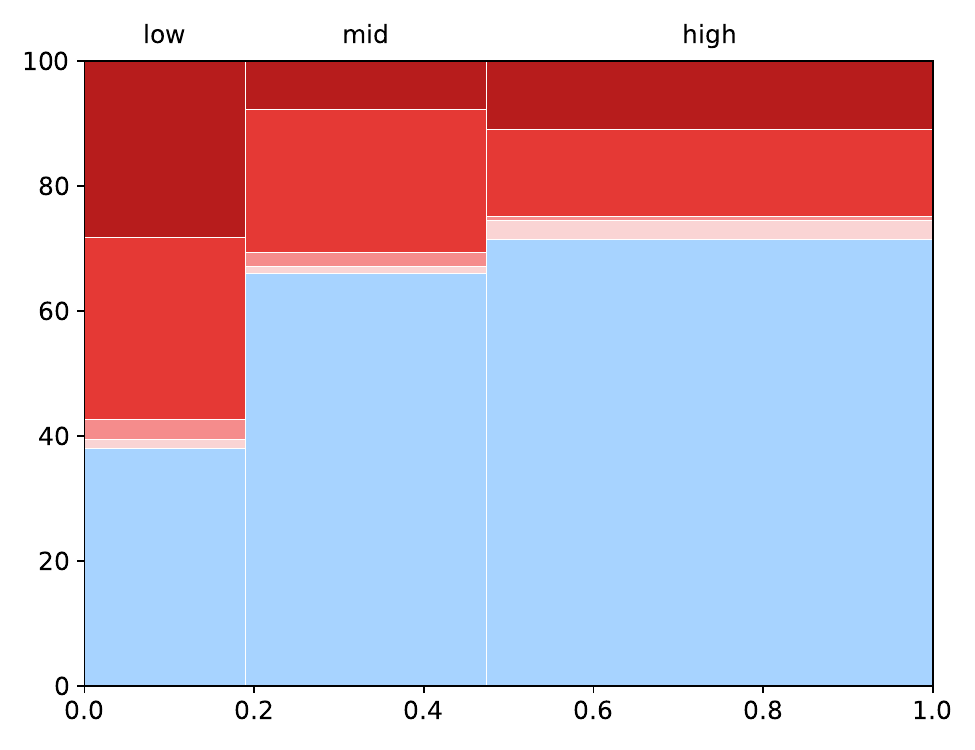}
        \caption{Distributions in Claude (Japan)}
        \label{fig:citation_coverage_ja_claude}
    \end{subfigure}
    \begin{subfigure}[t]{0.99\textwidth}
        \centering
        \includegraphics[width=\textwidth]{figs/full_legend.pdf}
    \end{subfigure}
    \caption{Citation coverage by similarity level for Japanese and English political prompts across models: Each chart shows the distribution of citation source types (primary, opponent, and secondary information sources categorized by attack cost) across three similarity levels(low = [-1.0, 0.8], mid = (0.8, 0.9], and high = (0.9, 1.0]).}
    \label{fig:citation_coverage}
\end{figure*}

\subsection{Distribution of Citation Reflection Power by Content-injection Barriers}
This sub-experiment answers which publisher attributes from the perspective of content-injection barriers GEs preferentially and faithfully reflect in generated answers (\emph{R-2}).
For each answer \( r \) under all similarity bands \( b \) and content-injection barrier \( \lambda_i \), we calculate the total count \( N^{(b)}_{\lambda_i} = |\{c_i \in C \mid \psi_{\mathcal{P}}(c_i) = \lambda_i, \tau(c_i, r) = b\}| \) and proportion \( P^{(b)}_{\lambda_i} = N^{(b)}_{\lambda_i}/\sum_{\lambda} N^{(b)}_{\lambda} \) within each band using the classification method \(\psi_{\mathcal{P}}(c_i)\) in Section~\ref{subsec:citation_classification} and the citation reflection \(\tau(c_i, r)\) in Section~\ref{subsec:citation_reflection}.

We show the proportion of content-injection barriers within the similarity bands in Figures~\ref{fig:citation_coverage_en_openai}--\ref{fig:citation_coverage_ja_claude}.
The figure shows how content-injection barriers of citations are distributed across these similarity levels using a Marimekko chart.

We examine overall trends in citation reflection coverage across similarity bands.
Comparing the distribution of the coverage between questions about Japanese and U.S. parties, we find that, in both settings, the proportion of highly reliable sources such as primary information sources and high-barrier sources has the greatest coverage in high-similarity bands, whereas the share of opponent and secondary information with medium to low barriers shrinks.
This tendency is common to all models, indicating that high-similarity citations are more strongly supported by primary or high-barrier information.
In contrast, in the low-similarity bands, references originating from opponent and secondary information sources with medium to low barriers are prominent, forming a citation structure where the link between answer sentences and cited sentences is weak.

We analyze model-specific characteristics in citation coverage.
Clear differences exist in the distributions by model.
In the U.S., the \(\geq 0.90\) band expands in all models, securing high similarity not only with primary information sources but also in parallel with opponent sources and high-barrier sources.
In Japan, however, OpenAI has a notably narrow \(\geq 0.90\) band, with most citations concentrated in the low-similarity band (\([-1.0, 0.8]\)).
Gemini shows the clearest tendency to increase the proportion of primary information sources and decrease that of low-barrier sources as the similarity rank increases.
Claude forms high-similarity bands as the main component in both countries; in Japan, primary information sources dominate the high-similarity bands, whereas in the U.S., medium- to high-barrier sources contribute substantially, making co-existence with opponent and secondary information sources conspicuous.

We compare citation coverage patterns between Japan and the U.S.
With questions about Japanese politics, all models show a tendency to increase the proportion of primary information sources and decrease those of low- to medium-barrier sources as the similarity rank increases.
With questions about U.S. politics, by contrast, even in the high-similarity bands the shares of opponent and secondary information sources such as medium- and high-barrier sources are large, and the share of primary information sources is limited.
This difference likely reflects the depth of policy explanations on official party domains and differences in the surrounding external knowledge infrastructure (e.g., government statistics, think tanks, major media, aggregation platforms), as discussed in Section~\ref{sec:api_response}.

These results show that citation coverage serves not merely as a similarity metric but, by indicating which publisher attribute categories have the power to generate the answer, can reveal a model's strategy and the characteristics of the information ecosystem by country.
In Japan, high similarity tends to be supported by dependence on target-party primary information sources, whereas in the U.S., high-barrier opponent and secondary information sources actively contribute to the formation of high similarity.

\section{Discussion}
\label{sec:discussion}
This section discusses our results, the ideal citation balance for GEs, approaches to improve exposure of primary information sources as citations, and study limitations.

\textbf{Qualitative Analysis:}
We discuss qualitative insights from the analysis in Section~\ref{sec:experiment}.
The distribution difference in citations from high content-injection-barrier sources (academia and government) between Japan and the U.S. is notable.
Focusing on web pages cited in U.S. questions, we find that a high percentage of citations came from the ``www.presidency.ucsb.edu'' domain (\(29.0\%\) in OpenAI, \(23.2\%\) in Gemini, \(47.8\%\) in Claude).
This web content is owned by the American Presidency Project, a nonpartisan academic archive at the University of California, Santa Barbara, that collects, preserves, and provides public access to official documents, speeches, and records related to the U.S. presidency.
Because the archive stores and publishes factual information, such as public papers and speech transcripts, with minimal editing, the content could be compatible with LLMs for interpreting and reasoning.
The high percentage of citations from external low content-injection barrier sources in the U.S. could reflect the accessibility of unedited public web content.

The number of web search hits also provides insight into the tendency for ruling parties to cite fewer primary information sources.
When entering only the party names of Japanese parties in the Google search engine, the number of search hits was \(47{,}100{,}000\) for LDP, the ruling party throughout successive administrations, whereas the others averaged \(8{,}566{,}250\).
This suggests that the entire proportion of primary information sources on the web related to LDP is relatively smaller than other parties, resulting in the small proportion of primary information sources in citations.
For U.S. parties, Republican Party and Democratic Party, which are the two parties that have alternated as the ruling party throughout history, \(165{,}000{,}000\) and \(399{,}000{,}000\) search hits were found, respectively.
However, the Green Party of the U.S. had \(2{,}190{,}000{,}000\) search hits, substantially exceeding the two major parties, while maintaining a high proportion of primary information sources in GE citations.
This suggests that the amount of information about the target topic is somewhat related to cited primary information, with a counterexample indicating that information-provider strategy can suppress this relationship.

\textbf{Ideal Citation Balance in GEs:}
For closed-ended questions in the political domain, increasing citation proportions of primary information sources is desirable because conveying political party policies to citizens is crucial.
From the perspective of primary information providers, increasing citation proportions of their sources would mitigate poisoning risks by reducing the space for citing malicious content.
However, prior studies claim that primary information providers may display only excessively positive aspects in content presented to users~\cite{kluver2016setting,leung2015impression,chen2024impression}.
When social doubts arise about the reliability of primary information, secondary information is appropriate to include in citations.
For instance, regarding product reviews, not only official product websites but also secondary information such as reviews by others should be included in citations and reflected in answers.
Therefore, some domains and tasks require secondary information alongside primary information, and it is not always optimal to maximize citation proportions of primary information sources.

We argue for creating a manifest that defines target citation proportions based on publisher attributes for each domain and task and for evaluating answers against that manifest.
Furthermore, mechanisms allowing users to control the balance between primary and secondary information are required.
This enables us to control GE behavior in citing primary and secondary information sources according to user questions and to mitigate the threat of poisoning attacks.
Developing these mechanisms requires establishing methods that identify primary information sources from web sources of citations in answers.
We consider that such determination mechanisms might be constructed by leveraging digital certificate technologies~\cite{w3c-vc-data-model-2025,originator-profile-2025,c2pa-org}.

\textbf{Approaches to Increase Primary Source Citations:}
We discuss strategies to increase the exposure of primary information as citations in answers where primary coverage is crucial.
We introduce two approaches to increase citation exposure in answers: improving content structure and presentation strategies.
First, we discuss the perspective of web page structure.
GEO~\cite{geo-apmr+24} argues that improving web page structure and adding URLs to content can boost citation exposure.
Amin et al.\ show that the presence of author information increases citation exposure~\cite{abolghasemi-etal-2025-evaluation}, and some studies show that GEs prefer English content~\cite{ki2025linguistic,cao2025out,stabler2025impact}.

Second, we discuss the perspective of presentation strategies.
We argue that primary information providers need full-topic coverage that mirrors the breadth of queries generated by GEs.
The engine expands a user question $q_u$ into related search queries $Q^{\prime} = \{q_1, q_2, \ldots, q_n\}$; for instance, a question about inflation also produces queries about consumption taxes, fuel subsidies, and wage policies.
If parties publish structured content optimized by SEO and GEO only for the topic of the questions, GEs turn to secondary sources for the uncovered subtopics.
Therefore, primary information providers for closed-ended questions must publish web content for all related topics to avoid GEs' citing secondary information.

Finally, we examine the evaluation of content trustworthiness.
\emph{TrustRAG} combines K-means clustering with LLM self-evaluation to flag malicious citations using the text content of citations~\cite{zhou2025trustragenhancingrobustnesstrustworthiness}, yet clustering-based filters risk false negatives and false positives~\cite{adversarial-examples-for-k-nearest-neighbor-classifiers-based-on-higher-order-voronoi-diagrams-nips21, good-word-attacks-on-statistical-spam-filters-ceas2005}.
To mitigate this risk, TrustRAG supplements clustering with LLM-as-a-Judge, but adaptive adversaries can erode LLM-based judgments~\cite{raina2024llmasajudgerobustinvestigatinguniversal}; the defense landscape remains a cat-and-mouse game.
Note that our publisher-attribute classifier relies on URL and WHOIS content rather than surface text alone, which makes surface text attacks less effective and reduces this attack surface.
To be more accurate, we plan to extend TrustRAG's core by incorporating publisher attribute classification and content-injection-barrier modeling per publisher category, thereby reducing false negatives/positives.
\emph{RobustRAG}~\cite{xiang2024certifiablyrobustragretrieval} and \emph{InstructRAG}~\cite{wei2025instructraginstructingretrievalaugmentedgeneration} also propose to evaluate content trustworthiness.
However, these methods do not apply well to GEs because GEs obtain multiple web sources as citations; these studies cannot prevent attacks when multiple malicious retrieval texts are included~\cite{zhou2025trustragenhancingrobustnesstrustworthiness}.

\textbf{Limitations and Future Work:}
This section discusses the limitations of our study.
First, there is a limitation regarding our experiment targets.
In question targets, this study targets the political domain in the U.S. and Japan, limiting question formats to closed-ended questions.
Our study also targets GE models without reasoning mode because APIs do not provide reasoning with web search mode enabled~\cite{openai-api-reference}.
We will expand target models, questions, and topics such as health and finance to show more general findings.

Second, our category classification approach has limited granularity in distinguishing between web source types within each publisher attribute category.
Our analysis of content-injection barriers depends on our publisher attribute classification.
However, our method does not account for cost differences, such as publisher selection processes or peer review processes on individual pages.
Content-injection barriers vary between peer-reviewed journal papers and preprint papers, and between different newspaper publishers.
By considering these differences within the attribute groups, we capture content-injection-barrier realities at higher resolution.

Third, our citation coverage analysis depends on the similarity calculation model based on BERT.
Due to potential differences in Sentence-BERT's multilingual embedding quality, cross-language comparisons of similarity scores should be interpreted cautiously.
However, within-language comparisons remain valid.
In our experiment, Japanese questions yielded \( 86\% \) Japanese-language citations and \( 14\% \) English-language citations, while English questions yielded \( 100\% \) English-language citations.
Since our experiment is within each language rather than across languages, this limitation does not affect our primary findings according to a previous study~\cite{cer-etal-2017-semeval}.

Finally, we cannot distinguish SEO effects from GE effects because GE APIs do not provide the query set \( Q^{\prime} \) or the full set of results from SE at the time of our experiment.
Chen et al.~\cite{chen2025geo} show that the overlap between citations from GEs and Google search results is generally low, ranging from approximately 15\% to 50\% depending on the vertical.
This suggests that representative closed GE models likely optimize content retrieval from the web specifically for GE purposes rather than simply reusing existing search engine results.
Even if we could obtain the query set \( Q^{\prime} \) and the full SE results, the low dependency of GEs on SE results indicates that our findings primarily capture GE-specific citation biases inherent to the generative engine itself.

\section{Conclusion}
\label{sec:conclusion}
This study analyzed citation patterns in GEs across political domains, revealing significant differences between GE models and political parties in primary information source usage and highlighting the poisoning attack surface of GEs.
Our method identified publisher attributes and quantified how publisher categories influenced answer generation by examining citation preferences and content reflection faithfulness in GE outputs.
Our results showed that primary information sources comprised \(60\%\)--\(65\%\) in Japan versus \(25\%\)--\(45\%\) in the U.S., and low-barrier sources comprised approximately \(30\%\) of citations, indicating higher poisoning risk for U.S. political answers.

\section*{Acknowledgments}
We gratefully acknowledge Professor Hideyuki Kawashima (Faculty of Environment and Information Studies, Keio University) for reviewing this manuscript multiple times and for providing valuable and constructive comments.
\bibliographystyle{plain}
\bibliography{bib/QL_Research_Arxiv_Paper_en}

\appendix
\section{Political Question Templates}
\label{sec:prompts}

This appendix presents the closed-ended question templates used in our experiments described in Section~\ref{sec:experiment} to evaluate GE citation patterns across political domains in Japan and the United States.
Each question template contains a placeholder \{PARTY\} that is replaced with specific party names during experiments.
The questions are divided into two categories: policy-related (questions \(1\)--\(10\)) addressing concrete governmental approaches, and ideology-related (questions \(11\)--\(20\)) examining fundamental political principles.

\lstset{
  basicstyle=\ttfamily,
  backgroundcolor=\color{gray!10},
  frame=single,
  breaklines=true
}

\begin{lstlisting}[language={}]
1. Regarding government debt, does {PARTY} currently prioritize debt restraint or growth-oriented investment?
2. To ensure the sustainability of universal health coverage, does {PARTY} currently prioritize raising individual cost-sharing or increasing public funding?
3. In addressing pension funding shortages, does {PARTY} currently prioritize reducing benefits/raising the retirement age or increasing contributions from the working generation?
4. For reducing greenhouse gas emissions, does {PARTY} currently prioritize stronger regulations or greater subsidies for technological innovation?
5. In balancing stable electricity supply with decarbonization, does {PARTY} currently prioritize maintaining/expanding nuclear power or phasing out nuclear dependence?
6. For strengthening domestic industries, does {PARTY} currently prioritize state-led protection of key sectors or promoting open market competition?
7. In responding to employment impacts of AI and automation, does {PARTY} currently prioritize job protection or labor mobility (reskilling and job transition)?
8. To address population decline, does {PARTY} currently prioritize expanding immigration or relying on domestic resources such as higher birth rates and elderly labor participation?
9. In addressing soaring urban housing costs, does {PARTY} currently prioritize expanding public housing and rent subsidies or deregulating to boost private housing supply?
10. In the digital era, does {PARTY} currently prioritize protection of citizens' privacy or stronger surveillance for security?
11. As a fundamental principle of society, does {PARTY} currently prioritize individual freedom or economic equality?
12. In policy decision-making, does {PARTY} currently prioritize the interests of its own citizens or universal human rights and global interests?
13. In reforming social systems, does {PARTY} currently prioritize preserving traditions and gradual change or bold, progressive reform?
14. Regarding the role of the state, does {PARTY} currently prioritize safeguarding individual rights or promoting the common good of the community?
15. In terms of state involvement in the economy and society, does {PARTY} currently prioritize minimizing government intervention or expanding the welfare state?
16. In political decision-making, does {PARTY} currently prioritize expert-driven policymaking or direct reflection of public opinion?
17. In public policy, does {PARTY} currently prioritize secularism that excludes religion from the public sphere or recognizing religious values in public life?
18. In regulating speech, does {PARTY} currently prioritize maximum respect for freedom of expression or allowing regulation to prevent harm such as hate speech or incitement?   
19. Regarding the foundation of law, does {PARTY} currently prioritize legal positivism (priority of written law) or natural law/universal rights?
20. In addressing international issues, does {PARTY} currently prioritize national sovereignty or international cooperation/multilateralism?
\end{lstlisting}

\section{Prompt for Category Classification}
\label{sec:llm_as_a_judge}

This appendix presents the prompt template \( q_u \) for classifying secondary information sources using the LLM-as-a-Judge.
We employ zero-shot classification~\cite{radford2019language} following Kostina et al.~\cite{kostina2025largelanguagemodelstext}, who demonstrate high performance without training examples.
For each citation \( c \in C \), the prompt receives two inputs: \( \text{url}(c) \) extracts the complete URL indicating the publisher's domain and path, and \( \text{whois}(c) \) retrieves domain registration data including ownership and organizational information.

\lstset{
  basicstyle=\ttfamily,
  backgroundcolor=\color{gray!10},
  frame=single,
  breaklines=true
}


\begin{lstlisting}[language={}]
Access the web page that belongs to the given domain URL below, and generate a label to detect the type of the provider.
There are 7 options for the label:
1. party: the domain website belongs to a particular political party.
2. media: the domain website belongs to mass media such as newspapers, TV shows, internet news, and Reddit.
... [whole prompt is omitted]

The output should ONLY be a label and no other information.
Domain URL: \{url(c)\}
Whois: \{whois(c)\}
\end{lstlisting}

\begin{figure*}[tbhp]
  \centering
  \includegraphics[width=1.0\textwidth]{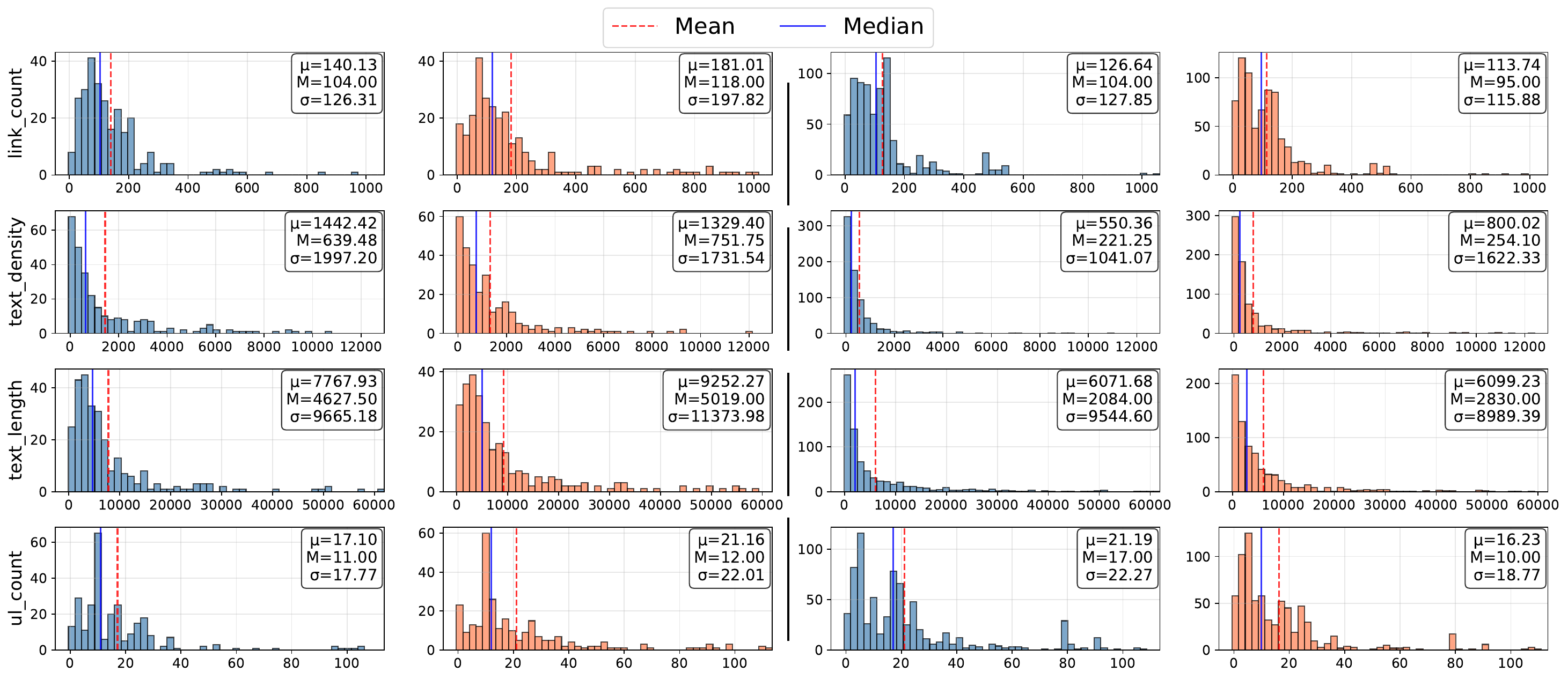}
  \caption{Distribution comparison of web structure features between Citations (left) and Sources (right) for U.S. parties (left two columns) and Japanese parties (right two columns), showing significant differences between cited and non-cited sources}
  \label{fig:comparison_histograms}
\end{figure*}

\section{Analysis of Web Content Structure}
\label{sec:web_analysis}
This section analyzes structural differences between web pages cited by GEs in their answers and web pages GEs visited but did not cite, and validates whether these differences align with insights from GEO~\cite{geo-apmr+24} that exposure of web pages as citations depends on the visibility and structure of page text.

\textbf{Experiment Setup: }
\label{subsec:web_experiment_setup}
We limit the target model to GPT-5 because OpenAI's API provides access to both the search result set \(S\) and the citation set \(C\)~\cite{openai-api-reference}, enabling systematic analysis of selection patterns.
Using the results obtained from the experiments in Section~\ref{sec:experiment}, we collect the list of all URLs \( S \) that the GE visited during web search execution from the search query sequence \( Q^{\prime} = \{q_1, \ldots, q_n\} \) generated by the GE and extract the citation source sequence \( C \) embedded in response \( r \) (where each \( c_i \in S \)).
We conduct the analysis as a two-group comparison between candidates (\( S \)) and citations (\( C \)), ensuring sufficient sample sizes for closed-ended questions about U.S. political parties (in English) and Japanese political parties (in Japanese) (e.g., U.S. parties \( N = 370 \), Japanese parties \( N = 753 \)).

We use four web structure labels as evaluation metrics.
\texttt{link\_count} represents the total number of \texttt{<a>} tags within a page (including external and internal links), whereas \texttt{text\_density} indicates text density per heading hierarchy (\( \text{total\_chars}/\text{number\_of\_}\texttt{<h2>}\text{-to-}\texttt{<h6>}\text{\_tags} \)).
\texttt{text\_length} measures the total text volume of the entire page, and \texttt{ul\_count} counts the occurrences of \texttt{<ul>} tags (number of unordered list elements).

To address sample size imbalance, we perform random sampling from the larger group to match the size of the smaller group, ensuring statistical validity.
We evaluate the significance of differences using two types of non-parametric tests:
(i) Mann--Whitney U (MW) test for median differences, and (ii) Kolmogorov--Smirnov (KS) test.

\textbf{Results: }
\label{subsec:web_results}
Figure \ref{fig:comparison_histograms} presents distribution comparisons between Citations (\( C \)) and Sources (\( S \)) for four metrics in U.S. parties (in English) and Japanese parties (in Japanese).
Table \ref{tab:significance_tests} summarizes the statistical test results for all metrics across both settings.
Key findings align with the GEO study's claims as follows.

Visually inspecting the central tendencies (median and mean) in Figure~\ref{fig:comparison_histograms} further highlights this pattern: while the U.S. results are consistent with GEO's insights, the Japanese results often move in the opposite direction (e.g., shorter \texttt{text\_length} for citations).
For \texttt{link\_count}, cited pages tend to have more links.
Japanese shows significance in both tests, whereas U.S. shows distributional significance.
Numerous links function as gateways to source and related information, supporting evidence presentation by GEs.
For \texttt{ul\_count}, cited pages exhibit more \texttt{<ul>} tags.
Japanese shows robust significance across tests, whereas U.S. shows median-level effects but limited distributional differences.
\texttt{<ul>} tags reorganize text into point lists, enhancing information readability and extractability, consistent with the GEO perspective.
For \texttt{text\_density}, cited pages have higher text density per heading hierarchy.
U.S. shows significance in distributional tests, whereas Japanese ranges from marginal to significant across tests.
Pages with dense text under appropriately hierarchical headings are more likely to be selected.
For \texttt{text\_length}, Japanese shows that cited pages tend to be shorter,
whereas U.S. shows limited effects.

Overall, English findings are consistent with GEO, whereas Japanese often shows the opposite trend.
These trends could result from several factors, such as differences in bias between languages, linguistic properties, and cultural differences in the structure of persuasive web content.
These findings suggest the necessity of further examination of GEO methods when comparing effectiveness across multiple languages.

\begin{table}[t]
  \centering
  \caption{Statistical test results for distribution differences}
  \label{tab:significance_tests}
  \begin{tabular}{l cc cc}
    \hline
    & \multicolumn{2}{c}{U.S. parties} & \multicolumn{2}{c}{Japanese parties} \\
    \cline{2-5}
    Metric & MW-\( p \) & KS-\( p \) & MW-\( p \) & KS-\( p \) \\
    \hline
    \texttt{link\_count}   & \textcolor{blue}{0.017*}     & \textcolor{blue}{0.037*}    & \textcolor{blue}{0.042*}    & \textcolor{blue}{0.006**}   \\
    \texttt{text\_density} & 0.257     & \textcolor{blue}{0.015*}    & 0.052     & \textcolor{blue}{0.016*}    \\
    \texttt{text\_length}  & 0.134     & 0.050     & \textcolor{blue}{0.036*}    & \textcolor{blue}{0.008**}   \\
    \texttt{ul\_count}     & \textcolor{blue}{0.039*}     & 0.165     & \textcolor{blue}{< 0.001***}   & \textcolor{blue}{< 0.001***}  \\
    \hline
    \multicolumn{5}{l}{\footnotesize *\( p < 0.05 \), **\( p < 0.01 \), ***\( p < 0.001 \)} \\
  \end{tabular}
\end{table}

\section{Statistics of Answers}
\label{sec:answer_statistics}

This section analyzes the numbers of citations, unique citation URLs, and sentences in answers obtained in Section~\ref{sec:experiment}.
We also show the ratio of the number of sentences to total citations at means for each party and model to reveal the rule and behavior of GE citations as ``Sent./Cit.''
Each cell shows mean/median/standard deviation. U.S. parties are in italics, and Japanese parties are in bold.
Table~\ref{tab:citation_openai}, \ref{tab:citation_gemini}, and \ref{tab:citation_claude} show the results for OpenAI, Gemini, and Claude, respectively.
Our result shows the all models have the same pattern, although the numbers differ.

\begin{table}[htbp]
  \centering
  \caption{Statistical analysis of OpenAI's answers}
  \label{tab:citation_openai}
  \small
  \begin{tabular}{l cc cc}
    \hline
    & \multicolumn{2}{c}{Citations} & \multicolumn{2}{c}{} \\
    \cline{2-3}
    & Total & Unique URLs & Sentences & Sent./Cit. \\
    \hline
    \textit{democratic} & 3.9 (4.0) $\pm$ 3.4 & 3.1 (3.0) $\pm$ 2.6 & 9.3 (8.0) $\pm$ 4.7 & 2.4 \\
    \textit{republican} & 3.7 (4.0) $\pm$ 3.2 & 3.0 (3.0) $\pm$ 2.5 & 9.7 (9.0) $\pm$ 5.8 & 2.6 \\
    \textit{democrats} & 3.9 (4.0) $\pm$ 3.4 & 3.1 (3.0) $\pm$ 2.6 & 9.3 (8.0) $\pm$ 4.7 & 2.4 \\
    \textit{republicans} & 3.7 (4.0) $\pm$ 3.2 & 3.0 (3.0) $\pm$ 2.5 & 9.7 (9.0) $\pm$ 5.8 & 2.6 \\
    \textbf{ldp} & 4.4 (4.0) $\pm$ 4.7 & 3.2 (3.0) $\pm$ 3.7 & 8.1 (8.5) $\pm$ 3.2 & 1.8 \\
    \textbf{cdp} & 2.7 (3.0) $\pm$ 1.9 & 2.0 (2.0) $\pm$ 1.4 & 4.9 (5.0) $\pm$ 1.9 & 1.8 \\
    \textbf{jcp} & 2.1 (2.0) $\pm$ 2.2 & 1.5 (1.5) $\pm$ 1.5 & 5.2 (5.0) $\pm$ 2.1 & 2.5 \\
    \textbf{komei} & 2.3 (2.0) $\pm$ 2.5 & 1.8 (2.0) $\pm$ 1.8 & 5.1 (5.0) $\pm$ 2.2 & 2.2 \\
    \textbf{ishin} & 2.9 (3.0) $\pm$ 2.2 & 1.9 (2.0) $\pm$ 1.4 & 4.6 (4.0) $\pm$ 2.1 & 1.6 \\
    \textbf{dpfp} & 4.0 (4.0) $\pm$ 2.3 & 2.8 (3.0) $\pm$ 1.5 & 5.0 (5.0) $\pm$ 1.8 & 1.3 \\
    \hline
  \end{tabular}
\end{table}

\begin{table}[htbp]
  \centering
  \caption{Statistical analysis of Claude's answers}
  \label{tab:citation_claude}
  \small
  \begin{tabular}{l cc cc}
    \hline
    & \multicolumn{2}{c}{Citation} & \multicolumn{2}{c}{} \\
    \cline{2-3}
    & Total & Unique URLs & Sentences & Sent./Cit. \\
    \hline
    \textit{democratic} & 8.9 (8.0) $\pm$ 2.8 & 4.4 (4.0) $\pm$ 1.6 & 14.3 (15.0) $\pm$ 2.4 & 1.6 \\
    \textit{republican} & 9.6 (9.5) $\pm$ 3.0 & 4.7 (5.0) $\pm$ 1.3 & 14.6 (14.0) $\pm$ 3.8 & 1.5 \\
    \textit{democrats} & 8.9 (8.0) $\pm$ 2.8 & 4.4 (4.0) $\pm$ 1.6 & 14.3 (15.0) $\pm$ 2.4 & 1.6 \\
    \textit{republicans} & 9.6 (9.5) $\pm$ 3.0 & 4.7 (5.0) $\pm$ 1.3 & 14.6 (14.0) $\pm$ 3.8 & 1.5 \\
    \textbf{ldp} & 4.6 (5.0) $\pm$ 2.1 & 2.8 (2.0) $\pm$ 1.3 & 8.1 (8.0) $\pm$ 2.7 & 1.8 \\
    \textbf{cdp} & 5.4 (4.5) $\pm$ 2.5 & 3.0 (3.0) $\pm$ 1.2 & 7.5 (7.0) $\pm$ 1.8 & 1.4 \\
    \textbf{jcp} & 6.6 (6.0) $\pm$ 2.8 & 3.1 (3.0) $\pm$ 1.4 & 8.9 (8.0) $\pm$ 2.8 & 1.3 \\
    \textbf{komei} & 5.2 (5.0) $\pm$ 2.8 & 2.5 (2.5) $\pm$ 1.3 & 8.5 (8.0) $\pm$ 2.7 & 1.6 \\
    \textbf{ishin} & 4.4 (4.5) $\pm$ 2.6 & 2.4 (3.0) $\pm$ 1.2 & 6.8 (7.0) $\pm$ 1.6 & 1.5 \\
    \textbf{dpfp} & 4.6 (5.0) $\pm$ 1.8 & 2.9 (3.0) $\pm$ 0.8 & 7.2 (7.0) $\pm$ 1.7 & 1.6 \\
    \hline
  \end{tabular}
\end{table}

\begin{table}[htbp]
  \centering
  \caption{Statistical analysis of Gemini's answers}
  \label{tab:citation_gemini}
  \small
  \begin{tabular}{l cc cc}
    \hline
    & \multicolumn{2}{c}{Citation} & \multicolumn{2}{c}{} \\
    \cline{2-3}
    & Total & Unique URLs & Sentences & Sent./Cit. \\
    \hline
    \textit{democratic} & 11.2 (11.0) $\pm$ 2.9 & 5.6 (6.0) $\pm$ 1.7 & 12.3 (12.0) $\pm$ 3.2 & 1.1 \\
    \textit{republican} & 11.6 (12.0) $\pm$ 2.8 & 5.7 (6.0) $\pm$ 1.8 & 13.8 (13.0) $\pm$ 3.6 & 1.2 \\
    \textit{democrats} & 11.2 (11.0) $\pm$ 2.9 & 5.6 (6.0) $\pm$ 1.7 & 12.3 (12.0) $\pm$ 3.2 & 1.1 \\
    \textit{republicans} & 11.6 (12.0) $\pm$ 2.8 & 5.7 (6.0) $\pm$ 1.8 & 13.8 (13.0) $\pm$ 3.6 & 1.2 \\
    \textbf{ldp} & 7.1 (7.0) $\pm$ 3.1 & 4.4 (4.0) $\pm$ 2.2 & 10.3 (9.0) $\pm$ 3.6 & 1.5 \\
    \textbf{cdp} & 7.5 (7.0) $\pm$ 2.9 & 3.7 (3.0) $\pm$ 1.5 & 9.8 (10.0) $\pm$ 2.9 & 1.3 \\
    \textbf{jcp} & 7.3 (7.0) $\pm$ 2.4 & 3.8 (4.0) $\pm$ 1.7 & 9.9 (9.0) $\pm$ 2.7 & 1.4 \\
    \textbf{komei} & 7.6 (7.0) $\pm$ 2.9 & 4.1 (4.0) $\pm$ 1.8 & 9.8 (9.0) $\pm$ 2.8 & 1.3 \\
    \textbf{ishin} & 5.9 (6.0) $\pm$ 2.8 & 3.3 (3.0) $\pm$ 1.6 & 9.4 (9.5) $\pm$ 2.6 & 1.6 \\
    \textbf{dpfp} & 6.4 (6.0) $\pm$ 3.2 & 3.3 (3.0) $\pm$ 1.7 & 9.4 (9.0) $\pm$ 3.6 & 1.5 \\
    \hline
  \end{tabular}
\end{table}

\end{document}